\documentclass[aps,prd,twocolumn,superscriptaddress,preprintnumbers,floatfix,nofootinbib,notitlepage,showkeys,showpacs]{revtex4}

\usepackage{graphicx}
\usepackage{graphics,psfrag}
\usepackage{latexsym}
\usepackage{amsmath,amssymb}
\usepackage{amsfonts}
\usepackage{enumitem}
\usepackage{slashed}

\newcommand{\ev}[1]{\left \langle #1 \right  \rangle}
\newcommand{\Dslash}{ \slashed D}
\newcommand{\be}{\begin{equation}}
\newcommand{\ee}{\end{equation}}
\newcommand{\bea}{\begin{eqnarray}} 
\newcommand{\eea}{\end{eqnarray}}

\begin{document}

\title{Walking Dynamics Guaranteed}
\author{Jarno Rantaharju}
\email{jarno.rantaharju@helsinki.fi}
\affiliation{Department of Physics \& Helsinki Institute of Physics, P.O. Box 64, FI-00014 University of Helsinki}
\author{Claudio Pica}
\email{pica@cp3-origins.net}
\affiliation{CP3 -Origins \& IMADA, University of Southern Denmark, Campusvej 55, 5230 Odense, Denmark}
\author{Francesco Sannino}
\email{sannino@cp3-origins.net}
\affiliation{CP3 -Origins, University of Southern Denmark, Campusvej 55, 5230 Odense, Denmark}
\affiliation{Theoretical Physics Department, CERN, Geneva, Switzerland}

\begin{abstract}
We report evidence for a continuous transition from an infrared conformal phase to a chirally broken one in four dimensions.
We study a model with two Dirac fermions in the adjoint representation of an SU(2) gauge interaction and a chirally symmetric four-fermion interaction.
At large four-fermion coupling, the model goes through a transition into a chirally broken phase and infrared conformality is lost.
We show strong evidence that this transition is continuous, which would guarantee walking dynamics within the scaling region in the chirally broken phase.
\end{abstract}


\preprint{HIP-2020-5/TH}

\maketitle

\section{Introduction}\label{introduction}

Critical phenomena are related to phase transitions between distinct phases of matter. Conformal field theories are the natural theoretical framework to classify and analyse the associated phase transitions \cite{Wilson:1973jj}.

A renown example of phase transitions is the  number-of-flavor-driven quantum phase transition from an IR fixed point to a non-conformal phase where chiral symmetry is broken~\cite{Miransky:1996pd}. Several scenarios have been considered for this type of phase transition ranging from  a Berezinskii--Kosterlitz--Thouless (BKT)-like phase transition~\cite{Kosterlitz:1974sm}, used for four dimensions in~\cite{Miransky:1984ef,Miransky:1996pd,Holdom:1988gs,Holdom:1988gr,
Cohen:1988sq,Appelquist:1996dq,Gies:2005as}, to a jumping (non-continuous) phase transition~\cite{Sannino:2012wy}. The discovery that higher-dimensional representations could be (near) conformal~\cite{Sannino:2004qp} for a small number of flavors led to the well-known conformal window phase diagram of~\cite{Dietrich:2006cm} that  guides lattice investigations~\cite{Pica:2017gcb}.  

A smooth quantum phase transition in four-dimensional gauge-fermion theories is also known as \emph{walking}~\cite{Holdom:1988gs,Holdom:1988gr}. It is expected to enhance the effect of bilinear fermion operators in models of dynamical electroweak symmetry breaking.

The study of this transition in gauge-fermion theories is limited by the discontinuous nature of the fermion number.
It is not possible to approach the transition in a continuous way.
Instead we consider a transition driven by 
a chirally symmetric four-fermion interaction
in an otherwise infrared conformal gauge model.
Unlike the flavor number, the four-fermion coupling is a continuous parameter that can be tuned to be arbitrarily close to the transition. 
If this transition is continuous, the model in the scaling region is walking.

The model is also expected to posses a varying anomalous dimension in the infrared conformal phase \cite{Fukano:2010yv}.
Because of these features models such as the one investigated here were termed  {\it Ideal Walking} \cite{Fukano:2010yv,Yamawaki:1996vr}.

We study two Dirac fermions coupled to the adjoint representation of an SU(2) gauge interaction and a four-fermion term.
In previous work~\cite{Rantaharju:2017eej}, we have determined the mass anomalous dimension at the infrared fixed point for four values of the four-fermion coupling.
As predicted \cite{Fukano:2010yv}, we found an anomalous dimension that increases monotonously with the four-fermion coupling.
We also found preliminary evidence for a continuous transition between the chirally broken and conformally symmetric phases.
A similar continuous transition was also recently discovered in a Higgs-Yukawa model with a chirally symmetric interaction \cite{Catterall:2020yoe}.

In this study we confirm the continuous nature of the phase transition. We use lattice sizes up to $L=24$ and three different lattice spacings to measure an order parameter of the transition, the chiral condensate, and to identify the scaling of the order parameter in the vicinity of the transition point. We then study the correlation length of the chiral condensate through the chiral susceptibility. We observe stable scaling dimensions at each value of the lattice spacing, which is consistent with a second point in the gauge versus four-fermion coupling plane.

The discovery of walking dynamics is supported by the simultaneous occurrence of a continuous conformal to chirally broken transition along with the observed increasing value of the fermion mass anomalous dimension nearing the phase transition from the conformal phase.

\section{The Model}\label{themodel}

We study the SU(2) gauge field theory with 2 Dirac fermion flavors transforming according to the adjoint representation of the gauge group augmented with a four-fermion term,
\begin{align}
 S = S_G + \bar\Psi \Dslash \Psi - \frac{y^2}{\Lambda_{\textrm{UV}}^2} \left [ \left( \bar\Psi\Psi \right)^2 - \left(\bar\Psi \gamma_5\tau^3\Psi\right )^2 \right ]. \label{NJL_term}
\end{align}
The four-fermion term preserves a U(1)$\times$U(1) subgroup of the full SU(4) chiral flavor symmetry. This gauged Nambu--Jona-Lasinio (gNJL) model has convenient characteristics~\cite{Rantaharju:2016jxy}. It can be simulated without a sign problem and the enhanced symmetry at $y=0$ protects the four-fermion coupling from additive renormalization.

It is notable that infrared conformality is not immediately broken by a nonzero four-fermion term.
Instead, the model is attracted to an IR fixed point below a critical value $y<y_c$.
In previous work we have found varying anomalous dimensions at different values of the coupling $y$ in the conformal phase.
Since an anomalous dimension is a unique property of a fixed point, this suggests that there is a line of infrared fixed points in the coupling space accessible when changing the gauge coupling $g$ and the four-fermion coupling $y$ \cite{Yamawaki:1996vr,Catterall:2007yx,DelDebbio:2008zf,Hietanen:2008mr,DelDebbio:2010hu,DelDebbio:2010hx,DelDebbio:2015byq,Rantaharju:2015cne}.

Two compatible scenarios are at play for the conformal to chiral symmetry breaking transition. In the first scenario the infrared fixed line is attractive for arbitrarily large values of the gauge coupling, from zero and up to a finite value of the four-fermion coupling. At $y>y_c$ the renormalization group flow is no longer attracted to the IR fixed line, but runs to infinity, breaking chiral symmetry. This scenario is also compatible with a jumping transition \cite{Sannino:2012wy}.

\begin{figure} \center
\includegraphics[height=0.6\linewidth,width=0.6\linewidth]{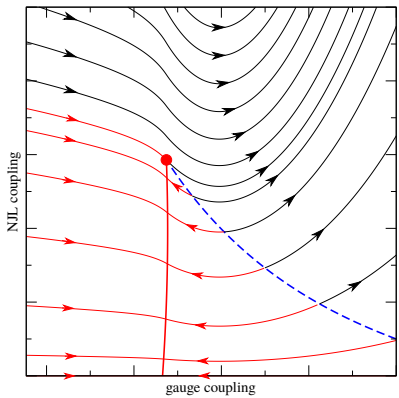}
\caption{A sketch of the fixed points and renormalization group flow in the fixed point merger scenario described in the text. The red line represents the infrared fixed line and the blue dashed line the conjectured ultraviolet critical line. The two lines merge at the red circle, creating a second order transition. }
\label{fp_rg_plot}
\end{figure}

The second scenario features a fixed point merger between the IR fixed line and a second line of corresponding to a UV fixed point at some larger value of the gauge coupling. As we increase the four-fermion coupling, the lines approach each other and merge at $y=y_c$. The merger induces a continuous transition \cite{Fukano:2010yv,Sannino:2012wy} point. This scenario is schematically represented in Figure~\ref{fp_rg_plot}.

To disentangle  a continuous transition from a jumping one, we study the order parameter of the chirally broken phase and the susceptibility of the chiral condensate on the lattice.

\subsection{Discretization}\label{discretization}

We start with the discretized version of the continuum action which reads
\begin{align}
 S &=  \beta_L \sum_{x,\mu<\nu} L_{x,\mu\nu}(U) + \frac{\sigma(x)^2+\pi_3(x)^2}{4a^2y^2} \\
 &+ \sum_{x} \bar\Psi(x) \left [ D_W + m_0+\sigma(x) + \pi_3(x) i \gamma_5\tau^3 \right ] \Psi(x)
  \label{action}
\end{align}
where $L_{x,\mu\nu}(U)$ is the plaquette discretization of the gauge action, $D_W$ is the Wilson Dirac operator and $a$ is the lattice spacing. The four-fermion term is enacted by employing two auxiliary fields, $\sigma(x)$ and $\pi_3(x)$.
The original action involving the four-fermion term is recovered by performing the integral of the partition function over the auxiliary fields.

\begin{figure} 
\includegraphics[height=0.6\linewidth]{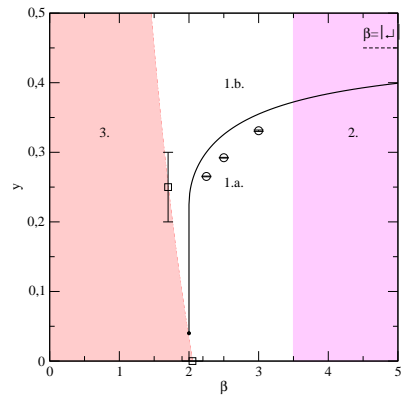}
\caption{ A sketch of the phase diagram at zero quark mass. Phases 1.a and 1.b are the physical infrared conformal and chirally broken regions. The solid line shows the large N ladder approximation result for the critical line separating the phases and circles denote its measured locations.
In region 2 at $\beta>\beta_{\textrm{max}}$, the Polyakov loop grows with $\beta$, indicating significant finite size effects. At $L=16$ we find $\beta_{\textrm{max}}\gtrsim 3$. In phase 3. there is a first order bulk transition instead of a critical line and the quark mass cannot be taken to zero. The squares denote the measured boundaries of this phase. }
\label{phase_plot}
\end{figure}

We show the zero mass phase diagram of the model in Fig. \ref{phase_plot}, including the new results on the location of the chiral symmetry breaking transition. A general description of the phase diagram can be found in \cite{Rantaharju:2017eej}. Here we concentrate on the edge of the chirally broken phase 1.b and avoid the unphysical finite size phase 2 and bulk phase 3. In phase 1.a the infrared dynamics of the model are dominated by the line of IR fixed points. In phase 1.b the infrared dynamics are described by a spontaneously broken chiral symmetry.

Due to the partially conserved chiral symmetry,
flavor multiplets are split into a single diagonal state and four non-diagonal states. Only the diagonal state corresponds to a symmetry. In the chirally broken phase the non-diagonal states gain an additional mass through the four fermion interaction and the spectrum includes a single massless Goldstone boson, the diagonal pseudoscalar meson.

This effect can be clarified by the Ward identities related to the axial flavour transformations,
\begin{align}
\partial_\mu &\ev{ A^{I,d}_\mu(x) O } = 2\bar m \ev{ P^d(x) O } \label{gsigma1} \\
&- 4 a^2 \bar y^2 \left( 1- \delta^{d,3} \right ) \ev{ S^0(x) P^d(x) O },\nonumber
\end{align}
Here $A_\mu^{I,d}$ is the axial current and $d$ labels the flavour symmetries, with $d=3$ labeling the diagonal direction. $P^d$ is the corresponding pseudoscalar density, $S^0$ is the singlet scalar density and $\bar y$ is a renormalized NJL coupling. The second term arises from the variation of the four fermion terms in the action and all order $1$ and $a$ terms have already been absorbed into a renormalized axial current. 
At the critical line the diagonal axial current is conserved and therefore $\bar m =0$. The second term will in general remain nonzero in the non-diagonal directions.

In the infrared conformal phase 1.a, two point functions decay according to a powerlaw and the first term must tend to zero at large distances. The last term must therefore also tend to zero in the infrared conformal phase.
In the chirally broken phase 1.b scale invariance is broken and the second term can become non-zero.
Replacing $O$ with $P^d(y)$ and setting $\bar m=0$ we find
\begin{align}
\partial_\mu \ev{ A^{I,d}_\mu(x) P^d(y) } \label{gbar_pcac} 
&=- 4a^2 \bar y^2  \ev{ S^0(x) P^d(x) P^d(y) }.
\end{align}
At large distances the right hand side approaches
\begin{align}
\partial_\mu \ev{ A^{I,d}_\mu(x) P^d(y) } \label{gbar_pcac} 
&= C \ev{P^d(x) P^d(y)}.
\end{align}

This is useful in defining an order parameter for the symmetry breaking transition,
\begin{align}
\bar m_{ND} = \frac{ \partial_0 \ev{ A^1_\mu(x) P^1(y) }}{ \ev{ P^1(x) P^1(y) } }.
\end{align}
At large enough separation $|x-y|$ this measures the breaking of the chiral symmetry.

Further, defining the chiral condensate as
\begin{align}
\Sigma_L = \frac 1V \sum_x\ev{S^0(x)},
\end{align}
we find at large $|x-y|$
\begin{align}
\partial_\mu \ev{ A^{I,d}_\mu(x) P^d(y) } \label{gbar_pcac} 
&=- 4a^2 \bar y^2  \ev{ S^0(x) P^d(x) P^d(y) } \\
&= -4a^2 C \bar y^2  \Sigma_L  \ev{ P^d(x) P^d(y) }, \nonumber
\end{align}
where $C$ is a constant.
This relation connects the chiral condensate to $\bar m_{ND}$ through a renormalized coupling $\bar y$. Since the coupling receives at most multiplicative renormalization, the quantity $m_{ND}$ provides a measurement of the chiral condensate up to a multiplicative renormalisation.

\subsection{Mass Scaling} \label{critical_mass}

The Wilson discretization of the fermion actions explicitly breaks the chiral symmetry, requiring the tuning of the mass counter-term $m_0$
at each pair of couplings $\beta$ and $y$. We set it following the methods described in reference~\cite{Rantaharju:2017eej}.
In the infrared conformal the masses of all states are zero when the 
fermion mass is zero. It is then straightforward to tune $m_0$ 
to the point where all measured meson masses reach zero.

\begin{figure} 
\includegraphics[height=0.6\linewidth]{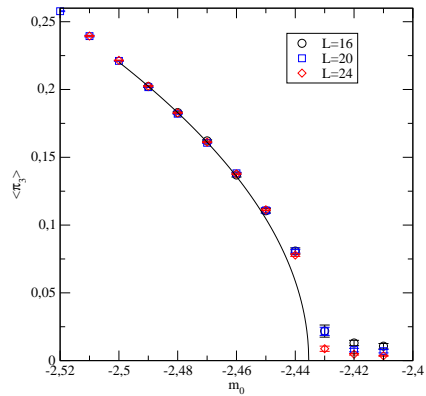}
\caption{ The tuning of $m_0$ in the chirally broken phase. The figure shows the expectation value $\ev{\pi_3}$ with $\beta=2.5$ and $y=0.3$ and the best fit to mean-field scaling. }
\label{pi_scaling_fig}
\end{figure}

In the chirally broken phase the value of $m_0$ is set by a second order transition in which the field $\pi_3$ acquires an expectation value and the associated correlation length diverges \cite{Rantaharju:2016jxy}.
The operator $\pi_3(x)$ sources the Goldstone of chiral symmetry breaking and the diverging correlation length is associated with the massless Goldstone state.
We therefore determine $m_0$ by measuring the condensate
\begin{align}
\ev{\pi_3} = \frac1V\ev{\sum_x\pi_3(x)}
\end{align}
as a function of the bare quark mass and fitting a second order scaling relation. An example of such a measurement is show in Figure~\ref{pi_scaling_fig}.

A full update consists of two HMC trajectories, the first only updating the auxiliary fields and the second updating both the gauge and the auxiliary fields. We use hypercubic lattices  with $V=L^4$ to measure auxiliary field expectation values and extended lattices with $V=2L\times L^3$ to measure meson masses and the order parameter for chiral symmetry breaking at the critical line.

The auxiliary field expectation values are measured
after each update and the autocorrelation is monitored for each parameter set. The autocorrelation grows when approaching the critical line and the chiral symmetry breaking transition as expected, but remains smaller that 10 full updates for most observables.
Close to the critical line the autocorrelation time of the average $\pi_3$ field increases and metastable states are observed during thermalization.

\begin{table*}
\center
\begin{tabular}{ c c | c c | c c }
$L$ & $\beta$ & $y$ & $m_0$  & $m_c$      & $\chi^2/d.o.f.$ \\ \hline
16 & 2.25 & 0.27  & -2.345  & -2.348(1) & 0.89 \\
   &      & 0.275 & -2.367  & -2.3659(8) & 0.25 \\
   &      & 0.28  & -2.391  & & \\
   &      & 0.29  & -2.435  & & \\
   &      & 0.3   & -2.478  & -2.4772(9) & 0.31 \\
   &      & 0.31  & -2.518  & & \\
   &      & 0.32  & -2.558  & -2.5579(4) & 0.70  \\
   &      & 0.33  & -2.594  & & \\
   &      & 0.35  & -2.661  & -2.6609(5) & 0.73 \\
   &      & 0.4   & -2.817  & & \\
   &      & 0.5   & -3.05   & -3.0496(2) & 0.78 \\ \hline
16 & 2.5  & 0.34  & -2.591 & -2.590(3) & 0.04  \\
   &      & 0.35  & -2.627 & -2.627(1) & 0.09 \\
   &      & 0.36  & -2.660 & -2.658(3) & 0.10 \\ \hline
16 & 3  & 0.35   & -2.577 & -2.577(4)  & 0.66 \\
   &    & 0.375  & -2.666 & -2.6660(5) & 0.67 \\
   &    & 0.39   & -2.715 &  &  \\
   &    & 0.4    & -2.745 & -2.745(1)  & 0.23 \\
   &    & 0.41   & -2.777 &  & \\
   &    & 0.42   & -2.806 &  & \\
   &    & 0.425  & -2.820 & -2.822(1)  & 0.82 \\
   &    & 0.43   & -2.835 &  & \\
   &    & 0.45   & -2.889 & -2.8893(8)  & 0.07 \\
20 & 2.25 & 0.28  & -2.393 & -2.3921(3) & 0.70 \\ \hline
   &      & 0.285 & -2.416 &  &  \\
   &      & 0.29  & -2.440 & -2.441(1)  & 0.20 \\
   &      & 0.295 & -2.459 &  &  \\
\end{tabular}
\quad
\begin{tabular}{ c c | c c | c c }
$L$ & $\beta$ & $y$ & $m_0$  & $m_c$    & $\chi^2/d.o.f.$ \\ \hline
20 & 2.25 & 0.3   & -2.480 & -2.4800(3) & 0.81 \\
   &      & 0.305 & -2.500 &  &  \\
   &      & 0.31  & -2.519 & -2.5182(3) & 0.36 \\ \hline
20 & 2.5  & 0.3   & -2.432 & -2.433(1)  & 1.0  \\
   &      & 0.31  & -2.476 & -2.4762(5) & 0.036\\
   &      & 0.315 & -2.495 &  &  \\
   &      & 0.32  & -2.515 & -2.5150(3) & 0.13 \\
   &      & 0.325 & -2.535 &  &  \\
   &      & 0.33  & -2.554 & -2.5542(2) & 0.12 \\
   &      & 0.335 & -2.573 &  &  \\ \hline
20 & 3    & 0.35  & -2.577 & -2.576(2)  & 0.12 \\
   &      & 0.355 & -2.595 &  &  \\
   &      & 0.3625& -2.622 & -2.6234(5) & 0.063\\
   &      & 0.37  & -2.648 &  &  \\
   &      & 0.375 & -2.665 & -2.6653(7) & 0.91 \\
   &      & 0.38  & -2.682 &  &  \\
   &      & 0.39  & -2.715 & -2.7150(5) & 0.56 \\ \hline
24 & 2.25 & 0.28  & -2.394 & -2.3930(3) & 0.76 \\
   &      & 0.275 & -2.370 & -2.369(1)  & 0.23 \\ 
   &      & 0.27  & -2.346 &    &  \\ \hline
24 & 2.5  & 0.298 & -2.424 & &  \\
   &      & 0.3   & -2.435 & -2.4326(3) & 2.9  \\
   &      & 0.305 & -2.454 & -2.454(2)  & 0.8  \\ \hline
24 & 3    & 0.34  & -2.539 &  &  \\
   &      & 0.345 & -2.558 &  &  \\
   &      & 0.35  & -2.577 & -2.578(1)  & 0.69 \\
   &      & 0.36  & -2.613 & -2.611(1)  & 0.75 \\
\end{tabular}
\caption{ Simulation parameters $L$, $\beta$, $y$ and $m_0$ used in the calculation of the order parameter $\bar m_{ND}$. The critical mass $m_c$ is recovered from the scaling fit Eq.~(\ref{pi_scaling}). The errors quoted include an estimate of systematic uncertainty found by varying the fit range. }
\label{parameters}
\end{table*}

The critical line is in the meanfield Universality class and follows the scaling relation.
\begin{align}
\ev{\pi_3} = C\left [m_0-m_c(\beta,y,L) \right]^{0.5}.\label{pi_scaling}
\end{align}
We find the critical mass $m_c(\beta,y,L)$ by fitting to this relation, setting the fit range to minimize $\chi^2/d.o.f.$
The results and $\chi^2$ values of the fits are given in Table~\ref{parameters}.
In addition we list the bare masses chosen for further simulations labeled as $m_0$.
Note that several values at $L=16$ from \cite{Rantaharju:2017eej} are included for comparison but not used in this study.

The determination of the critical line introduces a systematic uncertainty through the choice of the fit range. In the case shown in Fig.~\ref{pi_scaling_fig} the choice of fit range introduces a possible error of $\sim 0.005$ to the critical mass and similar uncertainties are found at different parameters. This translates to a systematic uncertainty of order $0.01$ in the measured order parameter $\bar m_{ND}$. We include this systematic uncertainty in reported values of $m_c$ and $\bar m_{ND}$.

\begin{table}
\center
\begin{tabular}{  c c c c c }
$\beta$ & $c_1$ & $c_2$ & $c_3$ & $\chi^2/d.o.f.$ \\ \hline
2.25 & -0.31(3) & -10.3(2) & 10.3(3) & 3.9 \\
2.5  & -0.36(8) & -9.5(5)  & 8.6(7) & 1.8 \\
3    & -0.64(5) & -7.4(3)  & 5.3(3) & 2.0 \\
\end{tabular}
\caption{The parametrization of the critical surface $m_0=m_c(\beta,y)$ in the chirally broken phase given in Eq.~(\ref{critical_line_fit}).  }
\label{mcrit_phasea}
\end{table}

In order to reduce the effect statistical and systematic uncertainties, we use a second order interpolating function to parametrize the critical line. At each value of $\beta$ we fit the measured values of the critical mass to
\begin{align}
m_c = c_1 + c_2y + c_3y^2
\label{critical_line_fit}
\end{align}
The values of the fit parameters and corresponding values of $\chi^2$ are given in Table \ref{table_mcrit_fit}.

While we are mainly interested in the chirally broken phase, we measure susceptibilities in the infrared conformal phase at $y<y_c$.
In order to determine the critical line in this phase, we measure the mass of the non-diagonal pseudoscalar meson. In an infrared conformal model this will scale as
\begin{align}
m_{PS} = C_{PS} \left [m-m_c(\beta,y,L) \right]^{\gamma}
\end{align}
At $\beta=2.25$ and $3$, we use the values found in the hyperscaling section of \cite{Rantaharju:2017eej}.
At $\beta=2.5$ we perform 5 measurements at 4 values of $y$ below $y_c$ with the lattice size $L=16$.
We find the location of the critical line $m_c$ by fitting to the relation above.

\begin{table}
\center
\begin{tabular}{ c c | c c | c c }
$L$ & $\beta$ & $y$ & $m_c$    & $\chi^2/d.o.f.$ \\ \hline
16 & 2.5  & 0.2   & -1.672(4) & 0.37  \\
   &      & 0.22  & -1.812(1) & 0.04  \\
   &      & 0.25  & -2.028(2) & 0.12  \\
   &      & 0.28  & -2.278(2) & 0.95  \\
16 & 2.25 & 0.25  & -2.196(3) & 2.3  \\
   &      & 0.2   & -1.828(2) & 0.76 \\
   &      & 0.1   & -1.357(5) & 1.1  \\
   &      & 0.05  & -1.241(8) & 1.73 \\

\end{tabular}
\caption{ Measured values of the location of the critical line in the infrared conformal phase, $y<y_c$.  }
\label{table_mcrit_fit}
\end{table}

\begin{table}
\center
\begin{tabular}{  c c c c c }
$\beta$ & $c_1$ & $c_2$ & $c_3$ & $\chi^2/d.o.f.$ \\ \hline
2.25 & -1.22(1)  &  0.42(2) & -17.2(5) & 0.9 \\
2.5  & -0.85(12) & -1.6(10) & -12(2)   & 2.2 
\end{tabular}
\caption{The parametrization of the critical surface $m_0=m_c(\beta,y)$ in the infrared conformal phase given in Eq.~(\ref{critical_line_fit}).  }
\label{table_mcrit_conformal_fit}
\end{table}

These results are included in table Table~\ref{parameters}. 
As in the chirally broken phase, we use a second order interpolation of the measured location of the critical line in the infrared conformal phase.
The coefficients are reported in Table~\ref{table_mcrit_conformal_fit}.

\section{Chiral Symmetry Breaking Transition} \label{transition}

\subsection{The Condensate}

Due to the partially broken chiral symmetry, it is straightforward to determine the chiral condensate without additive renormalization.
We use the partially conserved axial current (PCAC) relation
derived in \cite{Rantaharju:2016jxy}.
\begin{align}
\bar m_{ND}(t) &= -4a^2\bar y^2 \Sigma_L \\
&= \frac{ \sum_{\bf x} \partial_0 \ev{ A^{ND}_0(t,{ \bf x}) P^{ND}(0) }}{ \sum_{\bf x } \ev{ P^{ND}(t,{\bf x}) P^{ND}(0) } }.
\end{align}
Here $ND$ labels the non-diagonal directions of the flavor symmetry, which are broken by the four-fermion term.
In the chirally broken phase $\bar m_{ND}$ has a nonzero expectation value.

We stress that the existence of any dimensionful quantity breaks infrared conformality. While we expect the chiral symmetry to be broken, it is sufficient to find any nonzero dimensionful expectation value to show the existence of a phase transition.

\begin{figure*} \center
\includegraphics[height=0.3\linewidth]{{gsigma_fit_b2.25}.eps}
\includegraphics[height=0.3\linewidth]{{gsigma_fit_b2.5}.eps}
\includegraphics[height=0.3\linewidth]{{gsigma_fit_b3}.eps}
\caption{ The order parameter $\bar m_{ND}$ as a function of $y$ with $\beta=2.25$, $2.5$ and $3$ respectively. }
\label{gsigma}
\end{figure*}

\begin{table}
\center
\begin{tabular}{ c c | c c c }
$\beta$ & $\chi^2/d.o.f.$ & $y_c$ & $C$ & $\delta$   \\ \hline
2.25 & 0.33  & 0.2653(5) & 2.6(1)  & 0.69(2) \\
2.5  & 0.027 & 0.2918(3) & 2.59(7) & 0.72(3) \\
3    & 0.20  & 0.335(2)  & 1.9(2)  & 0.61(4) \\
\end{tabular}
\caption{ The scaling dimensions and coefficients of the order parameter $\bar m_{ND}$ in Eq.~(\ref{scaling_gsigma}). The errors include an estimate of the systematic error from varying the fit range. }
\label{scaling_table}
\end{table}

We measure the condensate $\bar m_{ND}$ along the critical line. At this phase we use an extended lattice with volume $V=2L\times L^3$. The parameters $y$, $m_0$ and $\beta$ are given in Table~\ref{parameters}.
When approaching the transition in the chirally broken phase, at leading order, we expect the condensate to scale as
\begin{align}
\bar m_{ND} = C_m \left ( y-y_c \right )^{\delta}. \label{scaling_gsigma}
\end{align}
Fits to this functional form are shown in Figure~\ref{gsigma} and the fit parameters are reported in Table~\ref{scaling_table}.
The figures also show values at $L=16$ from reference \cite{Rantaharju:2017eej} for comparison. These values are not used in the fits reported here.

The order parameter behaves as expected in a second order transition and shows little dependence on the lattice size.
The scaling dimensions are compatible within statistical and systematic accuracy, but the uncertainties are large.
The results point to a second order transition.

\subsection{Susceptibilities}

A second order transition is characterized by correlation lengths approaching infinity, leading to divergent susceptibilities.
As we readily see Figure \ref{gsigma}, the susceptibility of the order parameter with respect to the coupling $y$ diverges at the transition.
Directly measuring this susceptibility is unfortunately
complicated since it is generated by a four fermion term:
\begin{align}
\chi_{y} &= \frac{\partial}{\partial_y} \frac 1V \sum_{x} <S^0(x)> \\
= \frac 1V &\sum_{x,y} \ev{ S_0(x) \Pi(y) }- \frac 1{V^2} \sum_{x,y} \ev{S_0(x)} \ev{\Pi(y)},\\
\textrm{where } &\Pi(y) =S^0(y) S^0(y) - P^d(y)P^d(y)
\end{align}
This susceptibility is inconvenient to calculate from lattice data due to large statistical errors.

We may nevertheless derive some information on the transition by studying other susceptibilities. 
A convenient set can be defined as 
\begin{align}
\chi_{X} &= \frac 1V \sum_{x,y} \ev{ \bar\Psi(x)\Gamma_X\Psi(x) \bar\Psi(y)\Gamma_X\Psi(y) }\\
&- \sum_{x} \ev{ \right(\bar\Psi(x)\Gamma_X\Psi(x)\left)^2 }. \label{susceptibilities}
\end{align}
The $\Gamma_X$ is a matrix in Dirac and flavor space.
Here we measure three susceptibilities defined by
\begin{align}
\Gamma_1 = \tau^1 \textrm{, } \Gamma_{51} = \gamma_5\tau^1 
\textrm{ and } \Gamma_0 = I.
  \label{Gammas}
\end{align}
$\chi_0$ is the susceptibility of the chiral condensate with respect to the fermion mass.

\begin{figure*}
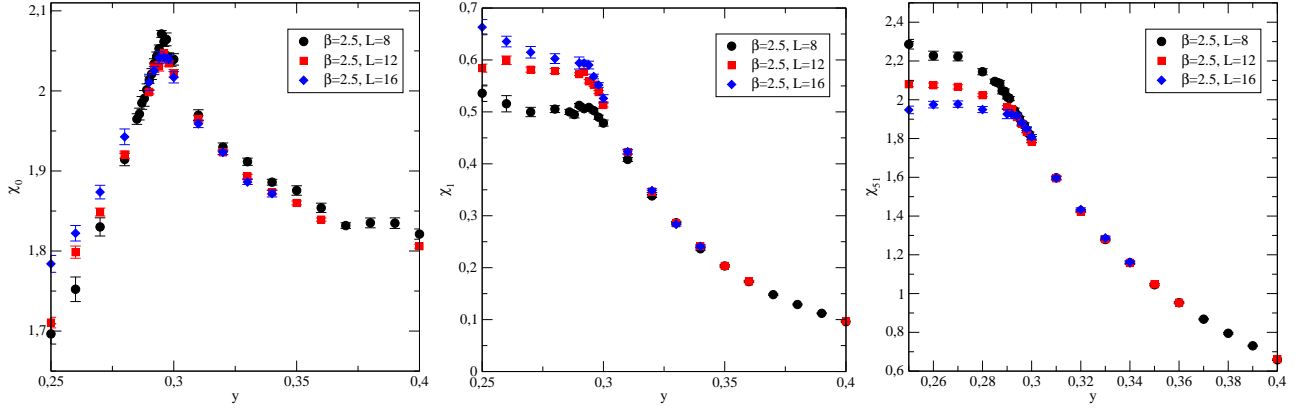

\includegraphics[height=0.3\linewidth]{{chi0_b2.5}.eps}
\includegraphics[height=0.3\linewidth]{{chi1_b2.5}.eps}
\includegraphics[height=0.3\linewidth]{{chi51_b2.5}.eps}
\caption{ Susceptabilities $\chi_{0}$, $\chi_{1}$ and $\chi_{51}$ respectively at $\beta=2.5$. }
\label{susceptibilities}
\end{figure*}

We show the behaviour of the susceptibilities at $\beta=2.5$ along the zero mass line at several values of $y$ and the volume is shown in Figure \ref{susceptibilities}.
Below the critical line susceptibilities $\chi_1$ and $\chi_{51}$ appear constant as a function of $y$, but large statistical errors prevent drawing a strong conclusion.
The chiral susceptibility $\chi_0$ decreases with $|y-y_c|$ and may be described by a powerlaw.

Above the critical line the measurements are more accurate and the susceptabilities are described by a powerlaw. 
The susceptibility of the chiral condensate, $\chi_0$,
behaves as
\begin{align}
\chi_\chi = C_\chi \left | y-y_c \right |^{-\nu}
\end{align}
Fits to the above relation are given in table \ref{susc_exp_table}.

\begin{table}
\center
\begin{tabular}{ c c c | c c c }
$\beta$ & L & $\chi^2/d.o.f.$ & $C_\chi$ & $y_c$ & $\nu$   \\ \hline
2.5 & 8  & 1.2 & 1.6(2) & 0.28(2) & 0.06(4)  \\
2.5 & 12 & 0.3 & 1.5(1)  & 0.28(2) & 0.09(3)  \\
2.5 & 16 & 0.8 & 1.49(6) & 0.28(1) & 0.08(2)   \\
2.25 & 8 & 1.3 & 1.74(5) & 0.27(1) & 0.04(1)  \\
2.25 & 12 & 0.2 & 1.69(3) & 0.263(3) & 0.045(5)  \\
2.25 & 16 & 0.4 & 1.71(2) & 0.265(2) & 0.039(5)  
\end{tabular}
\caption{ The powerlaw scaling of chiral condensate susceptibility $\chi_0$ above the critical point. }
\label{susc_exp_table}
\end{table}

\section{spectrum}

In this section we describe initial results for the meson spectrum of the model.
The spectrum can be classified according quantum numbers related to the global symmetries of the model.
In the absence of the NJL term, at $y=0$, the global flavor symmetry is SU(4).
Mesons transform according to the combinations $\bar{\textrm{4}}\otimes$4 which decomposes into the representations 1$\oplus$9. 
The NJL term explicitly breaks the flavor symmetry into U(1)$_\textrm{L}\times$U(1)$_\textrm{R}$.
This further devides the non-singlet representation into a conserved $U(1)$ and
8 degenerate states.
Since the unbroken generator contains $\tau^3$ in the space of the two flavors, we label it as the flavor diagonal subgroup.
Similarly we call the degenerate broken directions non-diagonal.
At $y > y_c$ the flavor symmetry is spontaneously broken into U(1)$_\textrm{V}$ and the coset U(1)$_\textrm{L}\times$U(1)$_\textrm{R}$/U(1)$_\textrm{V}$ contains a single flavor diagonal state with P$=-1$ and J$=0$. This is the Goldstone boson of the broken symmetry.

\begin{figure*}
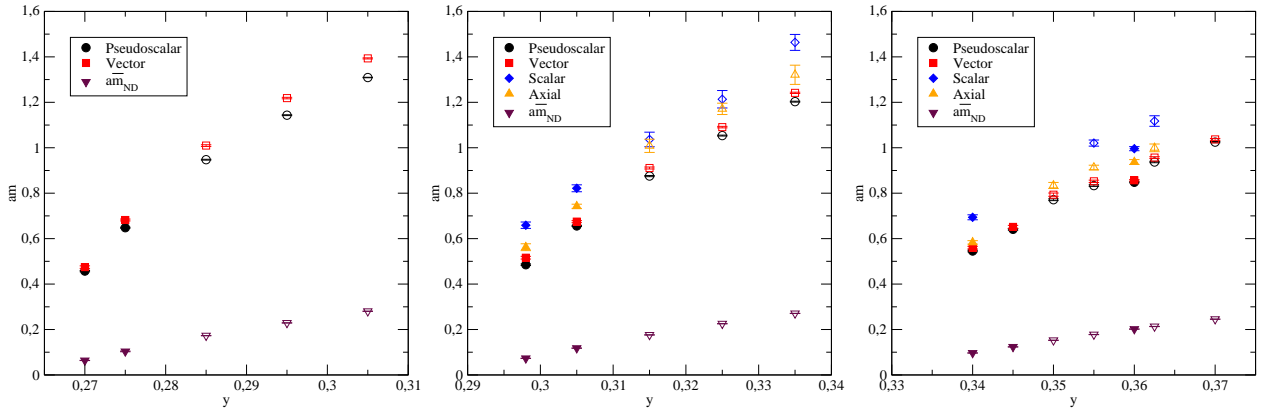

\includegraphics[height=0.3\linewidth]{{spectrum_b2.25}.eps}
\includegraphics[height=0.3\linewidth]{{spectrum_b2.5}.eps}
\includegraphics[height=0.3\linewidth]{{spectrum_b3}.eps}
\caption{ Mass spectrum of non-diagonal mesons $\beta=2.25$, $2.5$, and $3$ respectively as function of the condensate $m_{ND}$. 
Filled symbols denote lattice size $L=24$ and open symbols $L=20$. }
\label{spectrum}
\end{figure*}

Here we measure only the masses of the flavor non-diagonal states. Due to the auxiliary field terms in the fermion matrix, the propagators of flavor diagonal and flavor singlet states contain a disconnected contribution, making their measurement challenging.
We show the masses of non-diagonal pseudoscalar, vector and axial mesons (NDPS, NDV, NDA and NDS respectively) mesons as a function of the condensate in lattice units in Figure. \ref{spectrum}. The values of the non-diagonal meson masses and a selection interesting ratios are reported in table \ref{mesonmasstable}.

\begin{table*}
\center
\begin{tabular}{ c c c | c c c c c | c c c c c c }
$L$ & $\beta$ & $y$ & $\bar m_{\textrm{ND}}$ & $m_{\textrm{NDPS}}$ & $m_{\textrm{NDV}}$ & $m_{\textrm{NDA}}$ & $m_{\textrm{NDS}}$ & $\frac{m_{\textrm{NDV}}}{m_{\textrm{NDPS}}}$ & $\frac{m_{\textrm{NDV}}}{\bar m_{\textrm{ND}}}$ & $\frac{m_{\textrm{NDA}}}{m_{\textrm{NDPS}}}$ & $\frac{m_{\textrm{NDS}}}{m_{\textrm{NDPS}}}$ & $\frac{m_{\textrm{NDS}}}{\bar m_{\textrm{ND}}}$\\  \hline
24 & 2.25 & 0.27  & 0.0637(4) & 0.457(3) & 0.476(6) &         &         & 1.04(2)  & 7.4(1)  &         &         &         \\
   &      & 0.275 & 0.104(2)  & 0.648(2) & 0.682(4) &         &         & 1.052(7) & 6.6(1)  &         &         &         \\
20 & 2.25 & 0.285 & 0.1729(4) & 0.948(2) & 1.010(4) &         &         & 1.07(5)  & 5.84(3) &         &         &         \\
   &      & 0.295 & 0.2290(5) & 1.143(2) & 1.219(3) &         &         & 1.07(4)  & 5.32(2) &         &         &         \\
   &      & 0.305 & 0.2803(3) & 1.309(1) & 1.393(3) &         &         & 1.06(2)  & 4.97(1) &         &         &         \\\hline
24 & 2.5  & 0.298 & 0.0731(5) & 0.485(4) & 0.516(6) & 0.56(2) & 0.66(2) & 1.06(2)  & 7.1(1)  & 1.15(3) & 1.36(3) & 9.0(2)  \\
   &      & 0.305 & 0.1179(4) & 0.656(2) & 0.675(5) & 0.74(1) & 0.82(2) & 1.03(8)  & 5.73(4) & 1.13(2) & 1.25(3) & 7.0(1)  \\
20 & 2.5  & 0.315 & 0.1762(7) & 0.875(3) & 0.911(4) & 1.01(3) & 1.04(3) & 1.041(6) & 5.17(3) & 1.15(4) & 1.18(4) & 5.8(2)  \\
   &      & 0.325 & 0.2255(4) & 1.054(2) & 1.091(4) & 1.17(3) & 1.21(4) & 1.035(4) & 4.84(2) & 1.11(3) & 1.15(4) & 5.4(2)  \\
   &      & 0.335 & 0.2709(4) & 1.203(2) & 1.242(3) & 1.32(4) & 1.46(4) & 1.032(3) & 4.58(2) & 1.10(4) & 1.22(3) & 5.4(1)  \\\hline
24 & 3    & 0.34  & 0.0967(5) & 0.545(4) & 0.560(6) & 0.58(1) & 0.69(1) & 1.03(1)  & 5.20(5) & 1.07(2) & 1.27(2) & 7.1(1)  \\
   &      & 0.345 & 0.1236(6) & 0.641(4) & 0.652(7) &         &         & 1.023(8) & 4.79(3) &         &         &         \\
   &      & 0.36  & 0.2018(4) & 0.848(2) & 0.857(3) & 0.94(1) & 1.00(1) & 1.023(8) & 4.79(3) & 1.11(2) & 1.17(1) & 4.93(5) \\
20 & 3    & 0.35  & 0.1526(6) & 0.771(4) & 0.793(7) & 0.83(1) &         & 1.03(1)  & 5.20(5) & 1.08(2) &         &         \\
   &      & 0.355 & 0.1778(6) & 0.833(4) & 0.852(5) & 0.91(1) & 1.02(2) & 1.023(8) & 4.79(3) & 1.10(1) & 1.22(2) & 5.7(8)  \\
   &      & 0.3625& 0.2134(5) & 0.937(3) & 0.957(4) & 1.00(2) & 1.12(2) & 1.021(5) & 4.48(2) & 1.07(2) & 1.19(3) & 5.2(1)  \\
   &      & 0.37  & 0.2455(5) & 1.025(3) & 1.037(7) &         &         & 1.012(5) & 4.22(2) &         &         &         \\
\end{tabular}
\caption{ The value of the condensate and the masses of non-diagonal meson states. }
\label{mesonmasstable}
\end{table*}

At several values of $\beta$ and $y$ we do not measure $NDA$ and $NDS$ correlators with sufficient accuracy to calculate the meson mass. This happens mainly at $\beta=2.25$ but also in some measurements at $\beta=3$. At $\beta=3$ the correlators do not alway reach a plateau and the mass cannot be extracted.

The non-diagonal states gain a mass through spontaneous chiral symmetry breaking and the mass grows with $\bar m_{ND}$ as expected.
The increase appears almost linear in the range of masses measured, but approaches zero faster than linearly at smallest values.
Finite size effects are clearly visible at $\beta=3$, where
the measurements at $L=20$ are higher than at $L=24$.

In order to compare the spectrum between our three different lattice spacings,
we must set the scale as a function of $\beta$.
We have attempted to calculate the gradient flow scale, \cite{Luscher:2010iy,Borsanyi:2012zs} which is commonly used to set the scale in chirally broken gauge models.
In our model we find the observable $\ev{t^2E(t)}$ is not monotonous as a function of $t$ and does not uniquely define a physical scale.

\begin{figure}
\includegraphics[height=0.6\linewidth]{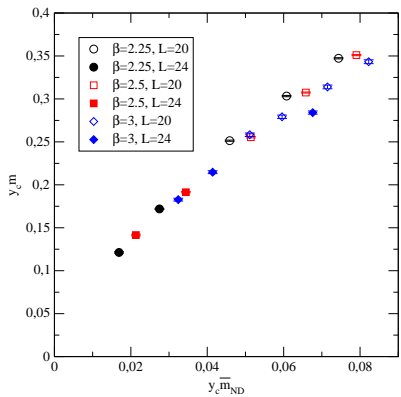}
\caption{ The mass of the non-diagonal pseudoscalar meson in units of $y_c$ with all three lattice spacings.
The spread of measurements at different lattice sizes and $\beta$ can be interpreted as finite size and lattice spacing effects. }
\label{pseudo_scaled}
\end{figure}

\begin{figure}
\includegraphics[height=0.6\linewidth]{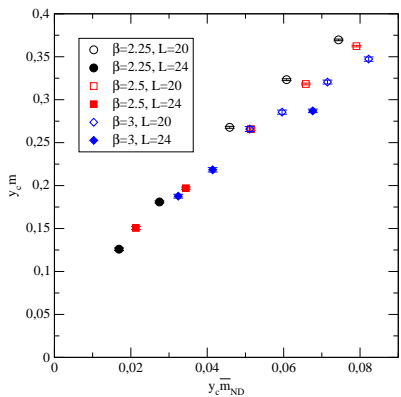}
\caption{ The mass of the non-diagonal vector meson in units of $y_c$ with all three lattice spacings.
The spread of measurements at different lattice sizes and $\beta$ can be interpreted as finite size and lattice spacing effects. }
\label{vector_scaled}
\end{figure}

Instead we use the critical NJL coupling to set the scale. The coupling has the dimension of length and does not receive additive renormalisation due to the enhanced symmetry at $y=0$.
The value of $y_c$ measured at each $\beta$ therefore defines a unique physical scale.
In Figures \ref{pseudo_scaled} and \ref{vector_scaled} we show the non-diagonal pseudoscalar and vector meson masses in units of $y_c$ as a function of the condensate $\bar m_{ND}$.

Rescaled with $y_c$, the masses of each state fall approximately on a line.
The width of the line can be intepreted as an measure of the discretisation and lattice size effects.
The effects are visible in each case, but significantly larger for the non-diagonal scalar and axial states.

\section{Conclusions} \label{conclusions}

We have investigated the quantum phase transition between the infrared conformal and the chirally broken phases of an SU(2) gauge model with two Dirac fermions in the adjoint representation and a chirally symmetric four-fermion term. We show compelling evidence for the discovery of a continuous transition between these two phases. To achieve our result, we measured the order parameter for the conformal phase for a range of four-fermion couplings $y$ and lattice spacings $\beta$. We found that the order parameter scales toward zero in agreement with a second order phase transition. We found a consistent scaling exponent for three different lattice spacings explored in this work.
These results are obtained on lattices large enough for our two coarser lattice spacings, to be considered at "infinite volume". 

We investigated the chiral susceptibility and found a finite value across the transition.
In the chirally broken phase the susceptibility follows the expected behavior dictated by a powerlaw decrease in the correlation length.
These results are compatible with a continuous transition.

The smooth nature of the newly unveiled four-dimensional quantum phase transition guarantees walking behavior in the scaling region at $y>y_c$. The model can be taken arbitrarily close to the infrared conformal phase below $y_c$ by adjusting the continuous parameter $y$. A smooth transition also guarantees the continuity of anomalous dimensions on both sides of the transition.

We also report on the mass spectrum of non-diagonal mesons.
We find that calculating the diagonal and scalar meson masses is numerically challenging due to disconnected contributions generated by the four fermion term.
The masses of non-diagonal mesons calculated in \cite{Rantaharju:2017eej} were compatible with an exponential scaling function.
Here we find significant dependence of the masses on the system size, especially at $\beta=3$.
This reduces our confidence on the earlier scaling results and we do not attempt to extract a scaling dimension using our new meson mass measurements.

For the future it is relevant to investigate the spectrum of the theory in the broken phase, paying special attention to the lightest scalar of the theory at large charges \cite{Orlando:2019skh}. The reason being that for a continuous quantum phase transition, a dilaton-like mode is expected to appear in the walking regime  to enforce approximate conformal invariance~\cite{Leung:1985sn,Bardeen:1985sm,Yamawaki:1985zg,Sannino:1999qe,Hong:2004td,Dietrich:2005jn,Appelquist:2010gy}.  
This subject has recently received renewed interest~\cite{Hong:2004td,Dietrich:2005jn,Goldberger:2008zz,Appelquist:2010gy,Hashimoto:2010nw,Matsuzaki:2013eva,Golterman:2016lsd,Hansen:2016fri,Golterman:2018mfm} due to recent lattice studies ~\cite{Appelquist:2016viq,Appelquist:2018yqe,Aoki:2014oha,Aoki:2016wnc,Fodor:2012ty,Fodor:2017nlp,Fodor:2019vmw} that reported evidence of the presence of a light singlet scalar particle in the spectrum.

\section{Acknowledgments}

This work was supported by the Danish National Research Foundation DNRF:90 grant, by a Lundbeck Foundation Fellowship grant and the U.S. Department of Energy, Office of Science, Nuclear Physics program under Award Number DE-FG02-05ER41368. Computing facilities were provided by the Danish Center for Scientific Computing and the DeIC national HPC center at SDU and the Extreme Science, Engineering Discovery Environment (XSEDE), which is supported by National Science Foundation grant number ACI-1053575, and CSC - IT Center for Science, Finland.

\end{document}